\documentclass[useAMS,usenatbib]{mn2e}

\usepackage{times}
\usepackage{amssymb}
\usepackage{rotating}
\usepackage{amsmath}
\usepackage{arydshln}

\input{epsf}

\title[QPO phase-resolved spectroscopy]{Phase-resolved spectroscopy of low frequency
  quasi-periodic oscillations in GRS 1915+105}
\author[A. Ingram \& M. van der Klis]{Adam
  Ingram$^{1}\thanks{E-mail:a.r.ingram@uva.nl}$ \& Michiel van der Klis$^{1}$\\
$^1$Astronomical Institute, Anton Pannekoek, University of Amsterdam, Science Park 904, 1098 XH Amsterdam, The Netherlands.\\}
\begin{document}

\date{Accepted 2014 November 6.  Received 2014 October 17; in original form 2014 August 28}

\pagerange{\pageref{firstpage}--\pageref{lastpage}} \pubyear{2014}

\maketitle

\label{firstpage}

\begin{abstract}
X-ray radiation from black hole binary (BHB) systems regularly
displays quasi-periodic oscillations (QPOs). In principle, a number of
suggested physical mechanisms can reproduce their power spectral
properties, thus more powerful diagnostics which preserve phase are
required to discern between different models. In this paper, we first
find for two \textit{Rossi X-ray Timing Explorer} (\textit{RXTE})
observations of the BHB GRS 1915+105 that the QPO has a well defined
average waveform. That is, the phase difference and amplitude ratios
between the first two harmonics vary tightly around a well defined
mean. This enables us to reconstruct QPO waveforms in each energy
channel, in order to constrain QPO phase-resolved spectra. We fit
these phase resolved spectra across 16 phases with a model including
Comptonisation and reflection (Gaussian and smeared edge components)
to find strong spectral pivoting and a modulation in the iron line
equivalent width. The latter indicates the observed reflection
fraction is changing throughout the QPO cycle. This points to a
geometric QPO origin, although we note that the data presented here do
not entirely rule out an alternative interpretation of variable disc
ionisation state. We also see tentative hints of modulations in the
iron line centroid and width which, although not statistically
significant, could result from a non-azimuthally symmetric QPO
mechanism.
\end{abstract}


\section{Introduction}
\label{sec:intro}

Low frequency quasi-periodic oscillations (hereafter QPOs), are
regularly observed in the X-ray light curves of accreting compact
objects in binary systems (e.g.\citealt{VDK2006}). Their properties
are tightly correlated with the observed spectral transitions in both
black hole and neutron star binaries (BHBs and NSBs). In BHBs, the QPO
fundamental frequency evolves from $\sim 0.1-30$ Hz as the spectrum
transitions from the power law dominated hard state to the
multicoloured disc blackbody dominated soft state
(e.g. \citealt{Wijnands1999a}). The multicoloured blackbody is well
understood as a geometrically thin, optically thick accretion disc
(\citealt{Shakura1973}; \citealt{Novikov1973}) and the power law as
Compton up-scattering of cool disc photons by energetic electrons in
some optically thin (optical depth $\tau \sim 1$) cloud near the black
hole (\citealt{Thorne1975}; \citealt{Sunyaev1979}). This cloud is
often interpreted as the evaporated inner accretion disc (\textit{the
  inner flow}; \citealt{Esin1997}; \citealt{Done2007};
\citealt{Gilfanov2010}), or alternatively the base of a jet
(\citealt{Markoff2005}; \citealt{Fabian2012}). The QPO signal
originates in the most part from this Comptonising cloud
(\citealt{Sobolewska2006}; \citealt{Axelsson2013}). When Comptonised
photons illuminate the disc, some fraction are scattered into the line
of sight with a characteristic reflection spectrum including a
prominent iron K$_\alpha$ line.

Suggested QPO mechanisms include relativistic precession models
(\citealt{Stella1998}; \citealt{Wagoner2001};
\citealt{Schnittman2006}; \citealt{Ingram2012}) and disc instability
models (e.g. \citealt{Tagger1999}; \citealt{Cabanac2010}). All of
these models can, in principle, reproduce the power spectral
properties of QPOs. Determining the QPO phase dependence of the
spectrum provides a powerful diagnostic tool to discern between
different models. This is relatively simple for periodic oscillations
such as eclipses and NS pulsations. In this case, phase-resolved
spectra can be constrained by folding the light curve
(e.g. \citealt{Gierlinski2002}; \citealt{Wilkinson2011}). However,
simply folding the light curve is not appropriate for QPOs, since
their phase does not evolve linearly, or even deterministically with
time (e.g. \citealt{Morgan1997}). In fact, it  is important to ask the
question: what makes QPOs \textit{quasi-periodic} rather than purely
periodic? Specifically, does the oscillation have an underlying
waveform, whereby the phase differences between different harmonics
and harmonic amplitude ratios are not random but instead have a well
defined average? We cannot tell using just the power spectrum and the
cross spectrum between energy bands if an oscillation with two strong
harmonics has some average underlying waveform or if it is simply
uncorrelated `noise' with a set of harmonically related characteristic
frequencies.

Here, in section \ref{sec:method} we show for two \textit{Rossi X-ray Timing
  Explorer} (\textit{RXTE}) observations of the BHB GRS 1915+105 that
the amplitude ratio and phase difference between the first two QPO
harmonics do indeed vary tightly around mean values (as is suggested
by measurements of the \textit{bicoherence} of the signal:
\citealt{Maccarone2011}). This indicates that there is some average
underlying waveform which, in sections \ref{sec:wave} and
\ref{sec:phasespec}, we estimate for each energy channel in order to
constrain QPO phase-resolved spectra. Then, in section
\ref{sec:spectral}, we fit these spectra with a model consisting of
Comptonisation and reflection to find strong spectral pivoting and a
modulation in the iron line equivalent width.

\section{The phase difference between QPO harmonics}
\label{sec:method}

Before we can phase-resolve the QPO, we must determine if there even
exists a well defined average underlying waveform. If the stochastic
process producing the QPO is instead uncorrelated between harmonic
frequencies, the meaning of phase resolved spectra is difficult to
asses. In this section, we show that there is indeed some average QPO
waveform by measuring the harmonic amplitudes and the phase difference
between harmonics.

\subsection{What makes QPOs quasi-periodic?}
\label{sec:decohere}

We can consider this question in general by representing the count
rate in the $k^{\rm th}$ time bin, $x_k$, as
\begin{equation}
x_k = \mu + \sum_{j=1}^{N/2} |X_j| \cos[ 2\pi j k / N - \varphi_j ],
\label{eqn:xk}
\end{equation}
where $X_j = |X_j| e^{i \varphi_j}$ is the discrete Fourier transform
(DFT) of $x_k$, $\mu$ is the mean count rate and there are $N$ time
bins in the light curve. Hereafter, we refer to $\varphi_j$ as the
\textit{phase offset} of the $j^{\rm th}$ Fourier frequency (which has a
frequency $\nu_j = j/[N~dt]$).

Splitting a long light curve which contains a QPO into many short
segments of length $N$ time bins allows us to study how the DFT at the
QPO harmonic frequencies varies between segments. Specifically, we can
measure how the amplitude and phase offsets vary with time for each
harmonic. If the oscillation was instead perfectly periodic, the
amplitude and phase offset of each harmonic would remain
constant. Since a QPO is only quasi-periodic, these conditions must
not all be met. Thus, perhaps a more insightful question is: how  does
the amplitude and phase offset vary for each QPO harmonic? Previous
work has already shown that the amplitude of the first two QPO
harmonics in XTE J1550-564 correlate with the flux over a $\sim 3$ s
timescale (\citealt{Heil2011}; also see \citealt{Ingram2011}). We note
that due to this, the \cite{Timmer1995} algorithm for generating
maximally stochastic time series is not appropriate for simulating
realistic QPO signals. In contrast, little is thus far known about how
the phase offsets vary.

Here, we consider variations in the phase offsets of the first two
QPO harmonics. Defining the phase offset of the $j^{\rm th}$ QPO
harmonic as $\Phi_j$ (as opposed to the phase offset of the $j^{\rm
  th}$ \textit{Fourier frequency}, $\varphi_j$), we can write
\begin{equation}
\Phi_2 = 2 ( \Phi_1 + \psi ),
\label{eqn:Phi2}
\end{equation}
where $\psi$ is the phase difference between the harmonics, defined on
the interval $0$ to $\pi$. From this definition, $\psi$ is the radians
of the first harmonic by which the $2^{\rm nd}$ harmonic lags the
$1^{\rm st}$. In general, $\psi$ will vary with time (i.e. from
segment to segment), but does it vary around a well defined mean value
or simply at random? In this paper, we study two observations of GRS
1915+105 which are described in the following subsection.

\begin{figure}
 \includegraphics[height=8.0cm,width=8.5cm,trim=1.5cm 0.0cm 0.0cm
 0.0cm,clip=true]{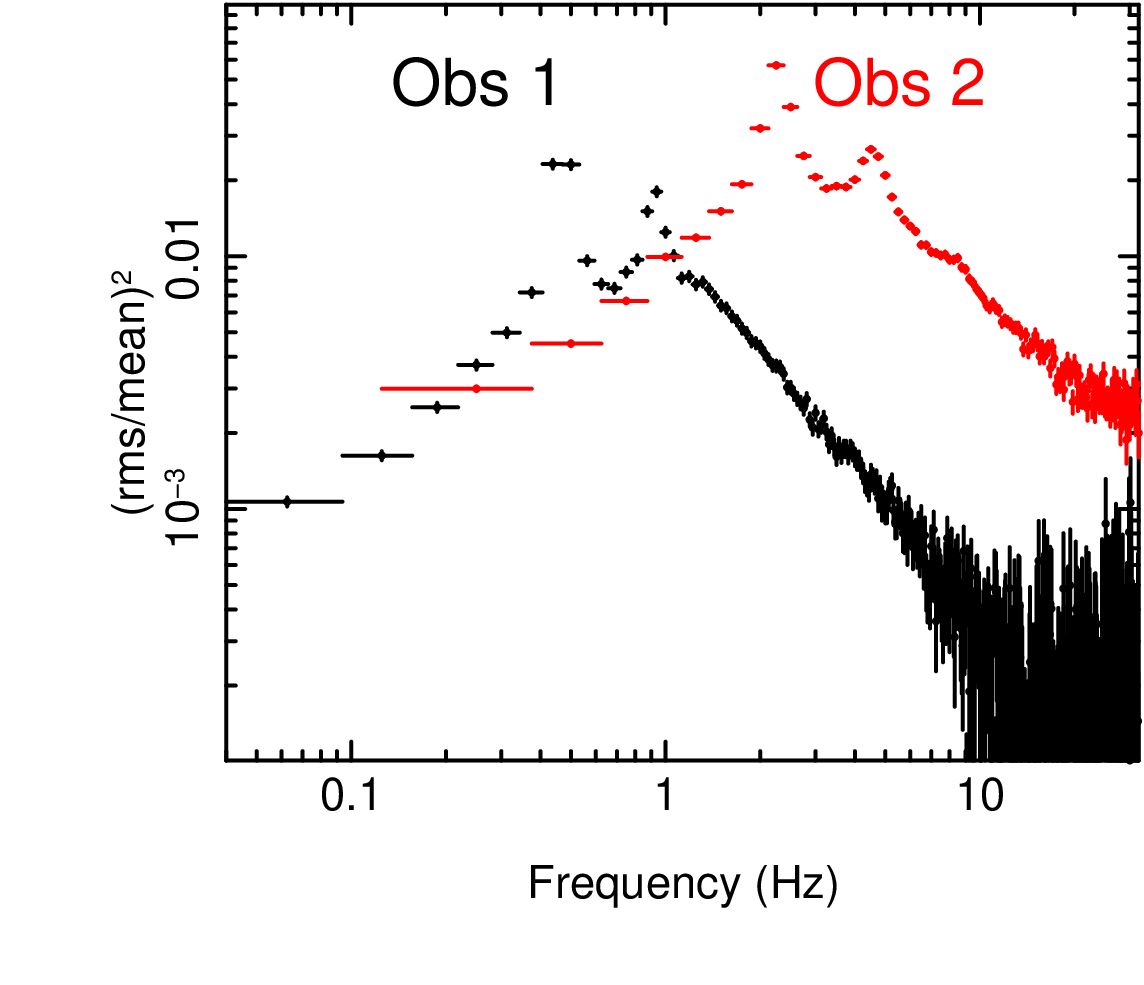}
\vspace{-10mm}
 \caption{White noise subtracted power spectra, plotted in units of
   frequency $\times$ power, for observations 1 (black) and 2
   (red). Both display strong QPOs with at least two harmonics.}
 \label{fig:psd}
\end{figure}

\subsection{Data}
\label{sec:data}

We consider two \textit{RXTE} observations of GRS 1915+105 with
observational IDs 60701-01-28-00 (hereafter \textit{observation 1})
and 20402-01-15-00 (hereafter \textit{observation 2}), both in the
$\chi$ variability class as defined by \cite{Belloni2000}. The white noise
subtracted power spectra of the full band light curves of both
observations are shown in Figure \ref{fig:psd}. Both clearly show QPOs
with strong contributions from the first two harmonics, which we fit
with Lorentzian functions in order to measure the centroid $\nu_0$ and
half width at half maximum (HWHM) for each component. For the
fundamental we measure $\nu_0 = 0.46$ Hz and HWHM$= 0.0275$ Hz for
observation 1 and $\nu_0 = 2.26$ Hz and HWHM$= 0.14$ Hz for observation
2. These two observations have been selected since they both have a
high count rate, a strong QPO and a (comparatively) long
exposure. Observation 1 was taken on March 6$^{\rm th}$ 2002 and
observation 2 was taken on February 9$^{\rm th}$ 1997. The time
averaged spectra for these observations can be well modelled by
an absorbed Comptonisation model with a soft power law ($\Gamma \sim
2.3$ and $\Gamma \sim 2$ for observations 1 and 2 respectively), in
addition to a strong contribution from a broad iron $K_\alpha$
emission line. Neither spectra require a direct disc component, but
this is simply due to the large absorption column around GRS 1915+105,
plus the hard response of the PCA. Whereas observation 2 was observed
to be radio faint (\citealt{Muno2001}), there are no radio data taken
simultaneous with observation 1 (\citealt{Prat2010}). However,
\cite{Yan2013} define two branches for GRS 1915+105 on a plot of
hardness ratio ($7-60$ keV flux / $2-7$ keV flux) against QPO
frequency and find that branch 1 and 2 approximately correspond
respectively to radio loud and quiet intervals (Figure 1
therein). Since observation 1 falls in branch 1, it is likely the
source was radio loud at this time.

We selected times when 3 and 5 Proportional Counter Units (PCUs) were
on for observations 1 and 2 respectively, in addition to applying
standard \textit{RXTE} good time selections (elevation greater than
$10$ degrees and offset less than $0.02$ degrees) using
\textsc{ftools} from the \textsc{heasoft} 6.15 package. After this
screening, observations 1 and 2 contain 9.680 ks and 10.272 ks of good
time respectively, and we measure mean count rates of 1744 c/s/PCU and
816 c/s/PCU. We extract light curves using \textit{saextrct}. Both
observations were taken in the `binned' mode ``B\_8ms\_16A\_0-35\_H'',
which has a timing resolution of $dt=1/128$ s and provides 16 energy
channels sensitive to the energy range $\sim 2-15$ keV. For the
purposes of spectral fitting, we generate response matrices using
\textsc{pcarsp} and background spectra using
\textsc{runpcabackest}. We also apply $0.5\%$ systematic errors using
\textsc{grppha} to account for uncertainties in the response of the
PCA and ignore the poorly calibrated lowest energy channel.

\subsection{Measuring the harmonic amplitudes and phase differences}
\label{sec:phase}

We split both light curves into $M$ segments, with each segment
containing $N$ time bins of duration $dt=1/128$ s. We may expect the
QPO to stay roughly coherent (i.e. periodic to a good approximation)
for $Q$ cycles, where $Q = \nu_0/{\rm FWHM}$ is the quality factor
(${\rm FWHM} = 2~{\rm HWHM}$). We therefore choose $N$ to ensure that
each segment contains $\sim Q$ cycles of the fundamental, whilst also
requiring $N$ to be an integer power of two in order to use the Fast
Fourier Transform algorithm. Thus $N \sim 1/({\rm FWHM}~dt)$, giving
$N=2048$, $M=605$ for observation 1 and $N=512$, $M=2548$ for
observation 2.

\begin{figure}
 \includegraphics[height=8.0cm,width=8.5cm,trim=0.0cm 0.0cm 0.0cm
 0.0cm,clip=true]{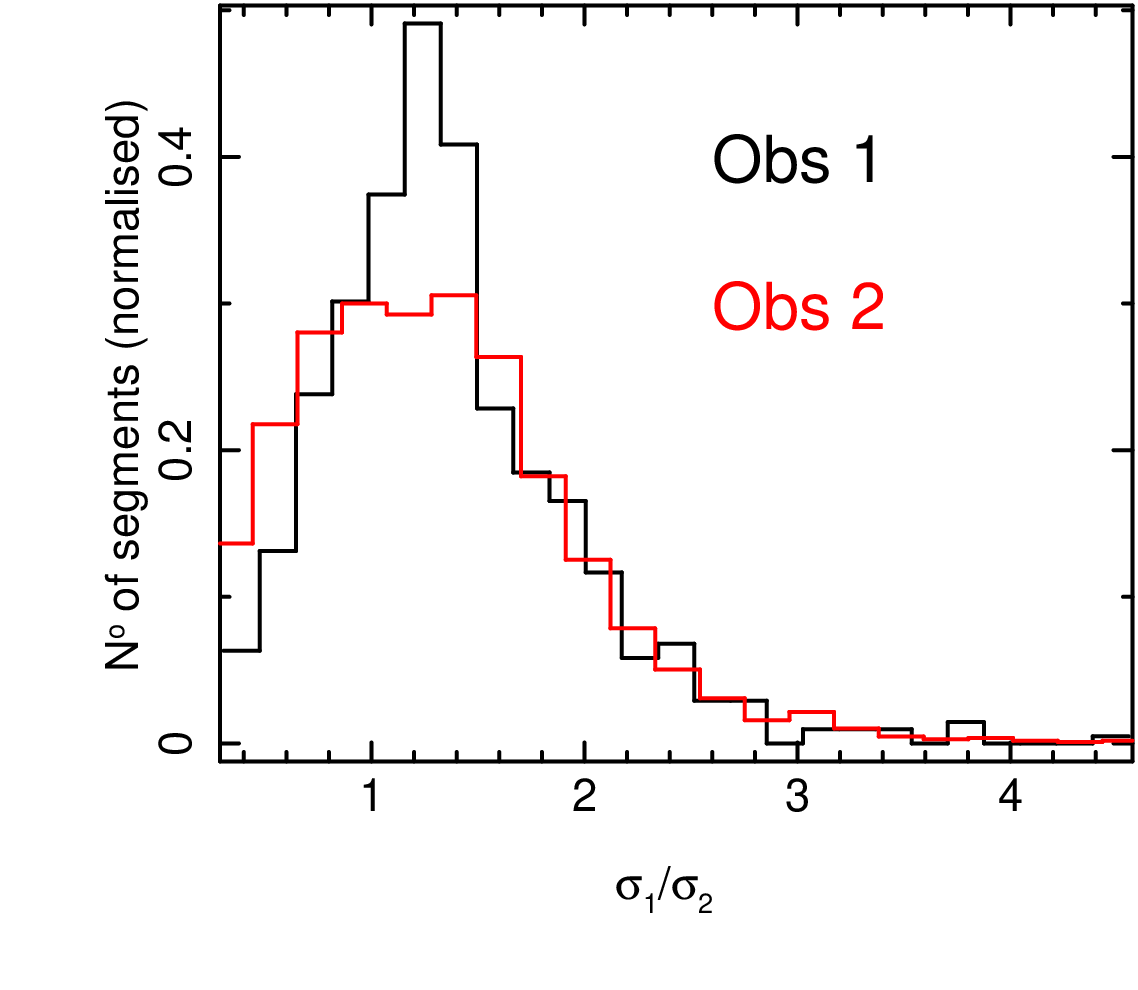}
\vspace{-10mm}
 \caption{Histogram of measured harmonic ratios for observations 1
   (black) and 2 (red).}
 \label{fig:sigdist}
\end{figure}

We first investigate the relative strength of each harmonic. We
measure the average rms in each harmonic using our multi-Lorentzian
fit to the power spectrum: the integral from zero to infinity of a
Lorentzian component gives the squared rms in that component. This
gives $\langle \sigma_1 \rangle = (9.6 \pm 0.2) \%$, $\langle \sigma_2
\rangle = (6.7 \pm 0.2) \%$ for observation 1 and $\langle \sigma_1
\rangle = (12.7 \pm 0.1) \%$, $\langle \sigma_2 \rangle = (7.6 \pm
0.1) \%$ for observation 2. We now wish to measure $\sigma_1$ and
$\sigma_2$ for each segment. Since the power spectrum calculated for
only one segment is very noisy, we cannot reliably fit a
multi-Lorentzian model for each segment. Instead, we use the centroids
and widths from our existing fit and calculate the power in the range
$\nu = \nu_0 \pm {\rm HWHM}$ for each segment. We then calculate the
normalisation of a Lorentzian function which has this integral in this
narrow range (i.e. similar to a bolometric correction). In Figure
\ref{fig:sigdist}, we plot a histogram of the harmonic ratio
calculated for each segment, $\sigma_1/\sigma_2$, normalised by the
number of segments in each observation. We see that, consistent with
previous work (\citealt{Heil2011}), the harmonic ratio appears to vary
around a well defined mean, although the distribution is more narrowly
peaked for observation 1.

If the phase difference between the harmonics, $\psi$, also has some
preferred value, we can conclude that the QPO does indeed have a well
defined mean waveform. For each segment, we calculate $\psi$ from the
phase offsets of the $1^{\rm st}$ and $2^{\rm nd}$ QPO harmonics,
$\Phi_1$ and $\Phi_2$ respectively, using the formula
\begin{equation}
\psi = [ \Phi_2/2 - \Phi_1 ]_{\rm mod \pi},
\label{eqn:psim}
\end{equation}
where the `${\rm mod \pi}$' signifies that each $\psi$
value is defined on the interval from $0$ to $\pi$. The phase offset
for the $j^{\rm th}$ Fourier frequency, which is simply the argument
of the DFT $X_j$, is given by
\begin{equation}
\tan\varphi_j = \frac{ \Im[X_j] }{ \Re[X_j] }.
\label{eqn:tanphi}
\end{equation}
We define $\Phi_j = \varphi_{jq}$, where $\nu_q$ is the nearest Fourier
frequency to the centroid of the fundamental. Since the width of each
Fourier frequency bin is $d\nu = 1/( N~dt )$, this means that $d\nu
\sim {\rm FWHM}$. Thus, by defining the QPO phase offsets in this way,
we are effectively averaging across the width of the fundamental.

\begin{figure}
 \includegraphics[height=8.0cm,width=8.5cm,trim=0.0cm 0.0cm 0.0cm
 0.0cm,clip=true]{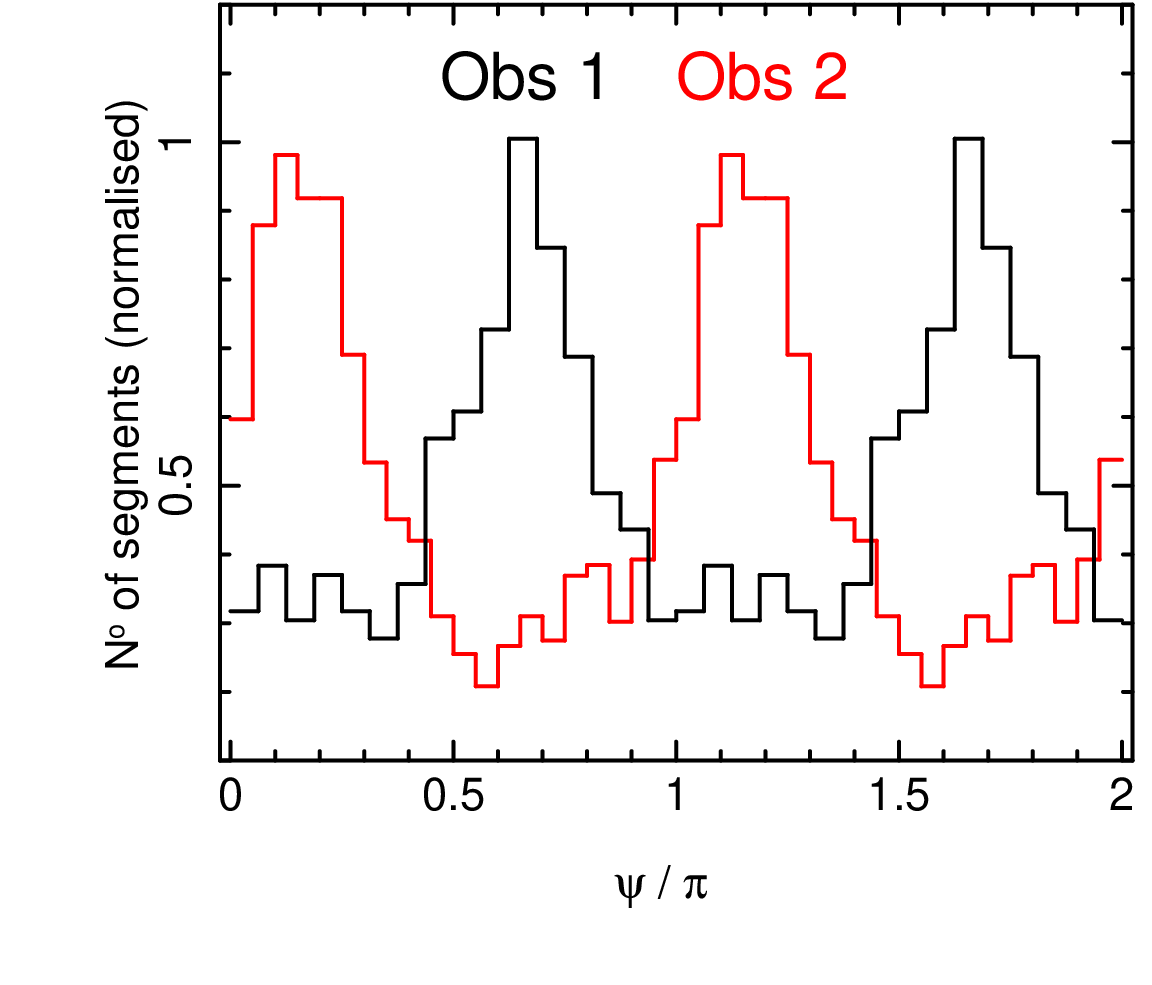}
\vspace{-10mm}
 \caption{Histogram of measured values of the phase difference between
   harmonics, $\psi$, for observations 1 (black) and 2 (red). The
   distributions each show two peaks purely because $\psi$ is defined
   on the interval $0\rightarrow\pi$ and we have shown two intervals
   by repeating the pattern.}
 \label{fig:dist}
\end{figure}

\begin{figure}
 \includegraphics[height=8.0cm,width=8.5cm,trim=0.0cm 0.0cm 0.0cm
 0.0cm,clip=true]{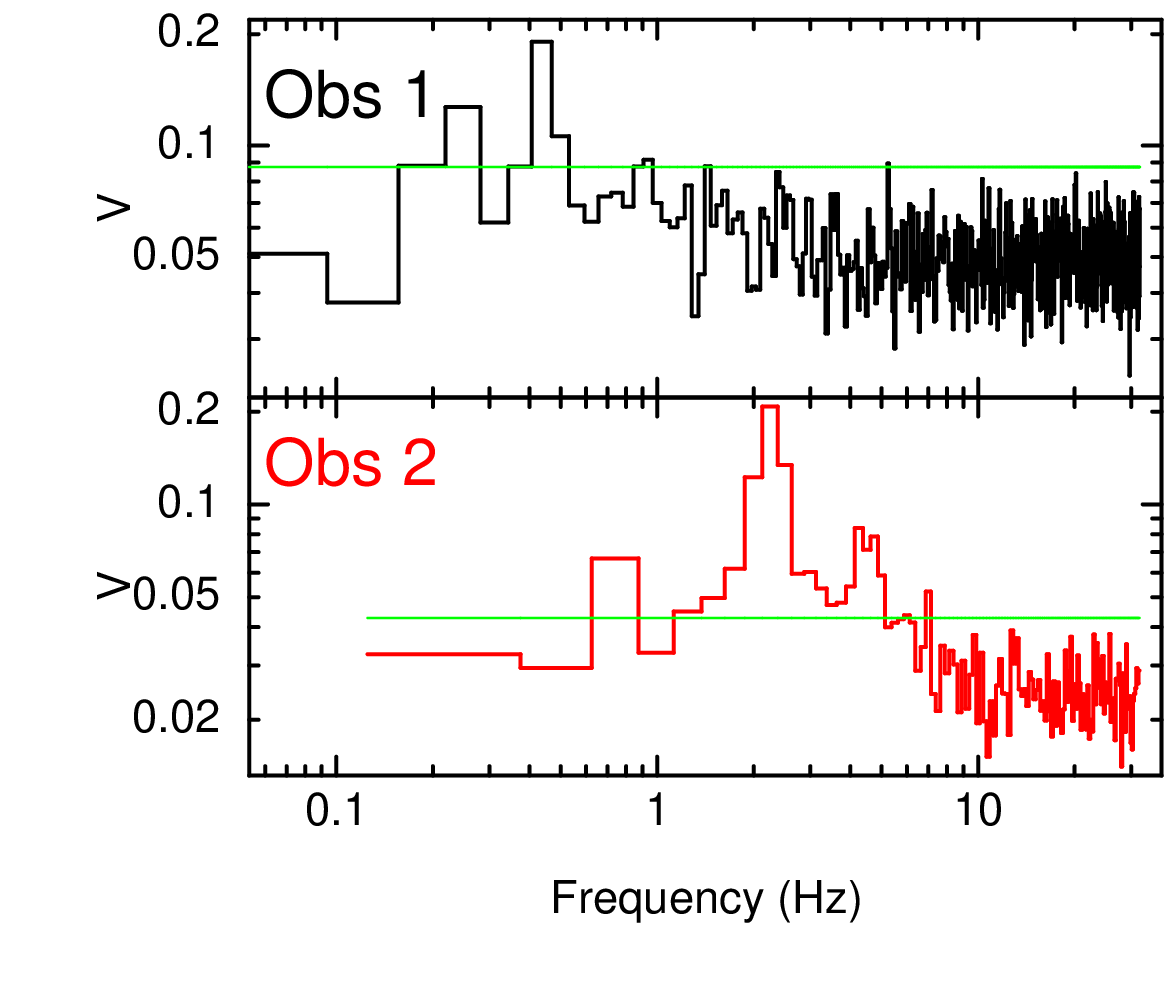}
\vspace{-10mm}
 \caption{
We measure the phase difference between the Fourier component at each
frequency, $\nu_j$, and twice that frequency, $2\nu_j$, for many
segments. $V$ is Kuiper's statistic, which assess the likelihood that
the phase differences are compatible with being drawn from a random
distribution. Values of $V$ above the green line are not compatible
with a random distribution ($>99.7\%$ confidence). Both observations
show a very strong peak at the fundamental QPO frequency, and the
higher signal-to-noise observation 2 also shows significant peaks at
higher harmonics (see text for details).} 
 \label{fig:KS}
\end{figure}

Figure \ref{fig:dist} shows a histogram of the $\psi$ values measured
for each segment. Here, we have defined phase bins on the interval $0
\leq \psi < \pi$ but we plot values up to $2\pi$ by repeating the
pattern. We see that for both observations, the phase difference
$\psi$ is clearly distributed around a mean value, indicating that
$\Phi_1$ and $\Phi_2$ do indeed correlate. Note that the distribution
only has one true peak: the second peak results because we have
repeated the pattern to account for the cyclical nature of
$\psi$. We formally confirm that the data are incompatible with a
random distribution using \textit{Kuiper's statistic} (see
e.g. \citealt{Press1992}). This is similar to a \textit{KS-test}, only
adapted to also be appropriate for a cyclical quantity such as the one
we are considering. It involves calculating the cumulative
distribution function of the data and measuring its maximum distance
above, $D_+$, and below, $D_-$, a theoretical cumulative distribution
function. The probability that the data belongs to the theoretical
distribution can be calculated from Kuiper's statistic, $V=D_+ + D_-$,
and the number of data points in the observed distribution. As
expected, this confirms the distributions shown in Figure
\ref{fig:dist} are not random with a significance $>> 5 \sigma$.

To compare the QPO with the broad band noise, we calculate Kuiper's
statistic for the series of phase differences between each Fourier
component ($\nu_j$) and the component with twice its frequency
($2\nu_j$). In Figure \ref{fig:KS} we plot $V$ against $\nu_j$ for all
$1 \leq j < N/4$ (since $2\nu_{N/4}$ is the Nyquist frequency). The
green lines indicate $3\sigma$ confidence intervals: if $V$ is above
the green line for a frequency $\nu_j$, the phase difference between
the components at $\nu_j$ and $2\nu_j$ are not randomly distributed
(at least with confidence $>99.7\%$). For both observations, all pairs
of broadband noise frequencies are consistent with a random phase
difference, in sharp contrast to the phase difference between $1^{\rm
  st}$ and $2^{\rm nd}$ QPO harmonics. We also see evidence of an
interaction between a sub-harmonic and the fundamental. Observation 2
additionally shows a deviation from  random phase differences between
the $2^{\rm nd}$ and $4^{\rm th}$ and even the $3^{\rm rd}$ and
$6^{\rm th}$ harmonics. This may provide a sensitive method for
detecting previously undetectable QPO harmonics. In this paper,
however, we concentrate on the interaction between $1^{\rm st}$ and
$2^{\rm nd}$ harmonics, which contain the bulk of the variability
power.

To measure the mean phase difference between $1^{\rm st}$ and $2^{\rm
  nd}$ QPO harmonics, $\langle \psi \rangle$, we must account for the
cyclical nature of $\psi$. For a particular trial value of $\langle
\psi \rangle$, the distance between $\psi_m$ and $\langle \psi
\rangle$ (in the $m^{\rm th}$ segment) is
\begin{equation}
d_m = 
\begin{cases} 
\delta          & \text{if $\delta<\pi/2$,}  \\
\pi - \delta & \text{otherwise,}
\end{cases}
\label{eqn:dm}
\end{equation}
where $\delta = | \psi_m - \langle \psi \rangle |$. This can be
understood by picturing all the $\psi_m$ values on a circle (with only
$\pi$ radians around its circumference): there are always two paths
around the circle to any point; $d_m$ is the shortest of these two
paths. We find the $\langle \psi \rangle$ value which minimises
$\chi^2 = \sum_{m=1}^{M} d_m^2$ using Brent's method
(e.g. \citealt{Press1992}), and calculate the standard deviation on the
mean as $\chi_{\rm min}/M$. This gives $\langle \psi \rangle / \pi =
0.667  \pm   9.9\times 10^{-3}$ and  $\langle \psi \rangle / \pi =
0.133   \pm   4.6\times 10^{-3}$ for observations 1 and 2
respectively. The fact that the phase difference is different between
observations indicates that the underlying QPO waveform has
changed.  In future, we will study how $\langle \psi \rangle$ depends
on QPO frequency for many observations.

\section{Reconstruction of the QPO waveform}
\label{sec:wave}

\begin{figure}
 \includegraphics[height=8.0cm,width=8.5cm,trim=0.0cm 0.0cm 0.0cm
 0.0cm,clip=true]{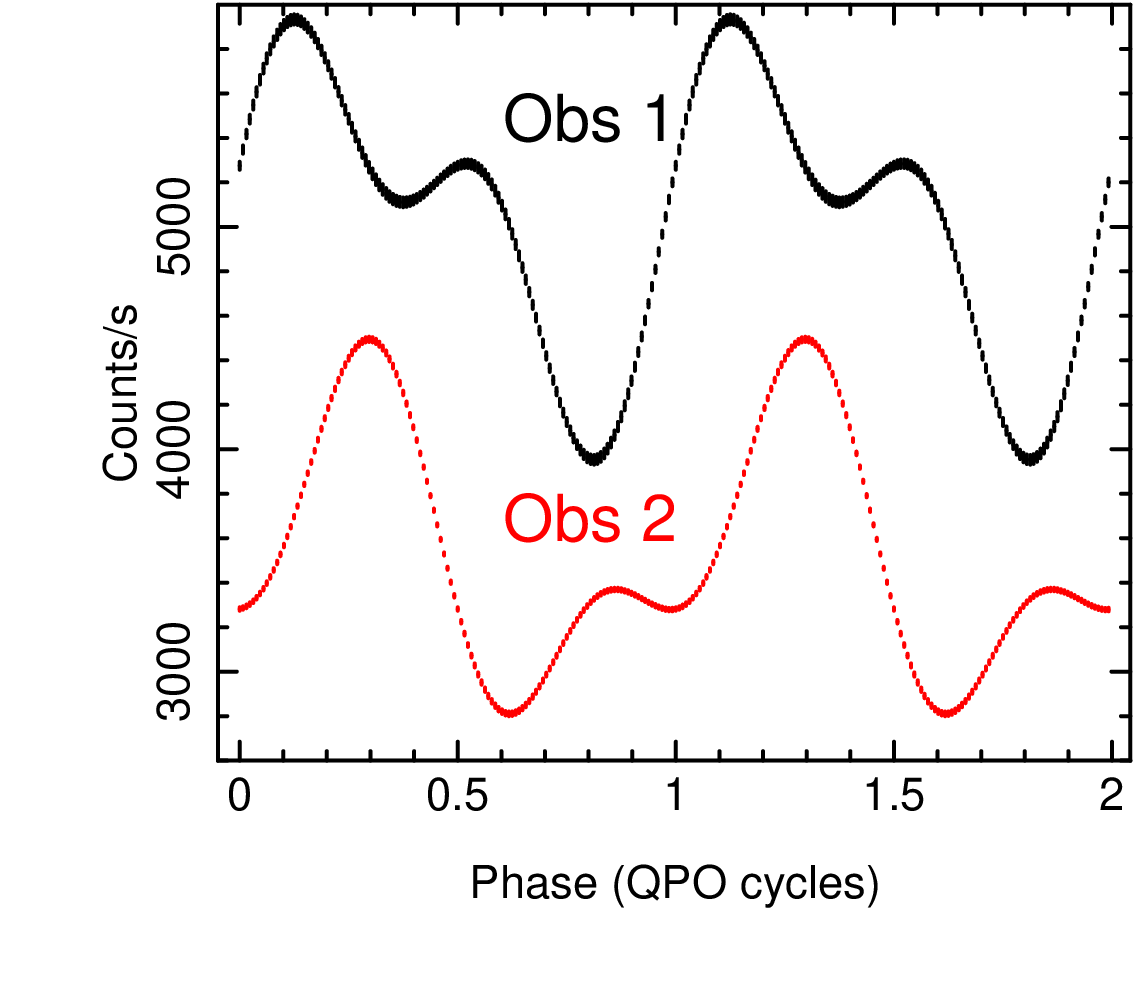} \vspace{-8mm}
 \caption{Reconstructed QPO waveform for both observations. The shape
   of the waveform differs dramatically between observations 1 (black)
   and 2 (red).}
 \label{fig:wave}
\end{figure}

Since we are able to measure average values for the amplitudes of, and
phase difference between, the first two QPO harmonics, we can
reconstruct an average underlying waveform. That is, we
can define a periodic function of QPO phase, $\phi$, given by
\begin{equation}
w(\phi) = \mu \left[ 1 + \sqrt{2}\sum_{j=1}^{J} \langle \sigma_j \rangle \cos( j\phi - \Phi_j )\right],
\label{eqn:w}
\end{equation}
where $\langle \sigma_j \rangle$ is the measured fractional rms in the $j^{\rm th}$
harmonic\footnote{the factor of $\sqrt{2}$ appears in equation
  \ref{eqn:w} because the variance of a sine wave is $1/\sqrt{2}$} and
$\Phi_j$ is the phase offset of the $j^{\rm th}$ harmonic. Here, the
phase offset of the first harmonic is arbitrary: we are interested in
the shape of the waveform rather than the starting point. We set
$\Phi_1=\pi/2$. The phase difference between each harmonic and the
first, in contrast, is important. Here we only consider $J=2$
harmonics since these contain the bulk of the power.

It is simple to measure the mean count rate $\mu$, and we use our
measurement of the phase difference between the first two harmonics,
$\langle \psi \rangle$, from the previous section. We also use our
measurements of $\langle \sigma_1 \rangle$ and $\langle \sigma_2
\rangle$ from the previous section. We then use equation \ref{eqn:w}
to obtain an estimate of the average underlying  QPO waveform. How
exactly this relates to the physical QPO mechanism depends, in
general, on the details of the processes generating the waveform and
those de-cohering it (\citealt{Ingram2013}). If the decohering process
is highly non-linear, this may introduce a bias in our estimate of the
true underlying waveform, or indeed may make such a true waveform
difficult to define. In the absence of a full understanding of all the
processes de-cohering the QPO, we define our waveform as a periodic
function with the average QPO properties.

To obtain an error estimate, we use our measurements of $\sigma_j$ and
$\psi$ for each segment (see the previous section), along with a
measurement of the mean count rate for each segment, in order to
calculate a waveform (using equation \ref{eqn:w}) for each
segment. This gives $M$ functions $w_m(\phi)$, in addition to our
estimate for the average waveform, $w(\phi)$. For each discrete value
of $\phi$ considered, we calculate the standard deviation on the mean
of the $w_m(\phi)$ points around the  average $w(\phi)$. Figure
\ref{fig:wave} shows the resulting waveforms for both observations,
evaluated for 128 QPO phases. Our reconstructed waveforms have small
errors since we can measure each of the five parameters in equation
\ref{eqn:w} accurately, and there are correlations between these
parameters. Note that different phase values in Figure \ref{fig:wave}
are not statistically independent of one another, and so we do not
expect to see a scatter in the data consistent with the size of the
error bars, nor will we be able to reduce the size of the errors by
binning on phase. The errors determined here are errors on the
\textit{function} $w(\phi)$ and the phase values are
\textit{instances} rather than intervals.

We note that this is not the first derivation of a QPO
waveform. \cite{Tomsick2001} used a folding method to estimate
the QPO waveform in observations of GRS 1915+105. This method, as
expected, yields similar results to ours but crucially, it implicitly
assumes that the phase difference between QPO harmonics is
\textit{constant}, which we find to only be approximately true.

\section{Phase resolving method}
\label{sec:phasespec}

\begin{figure}
 \includegraphics[height=8.0cm,width=8.5cm,trim=0.0cm 0.0cm 0.0cm
 0.0cm,clip=true]{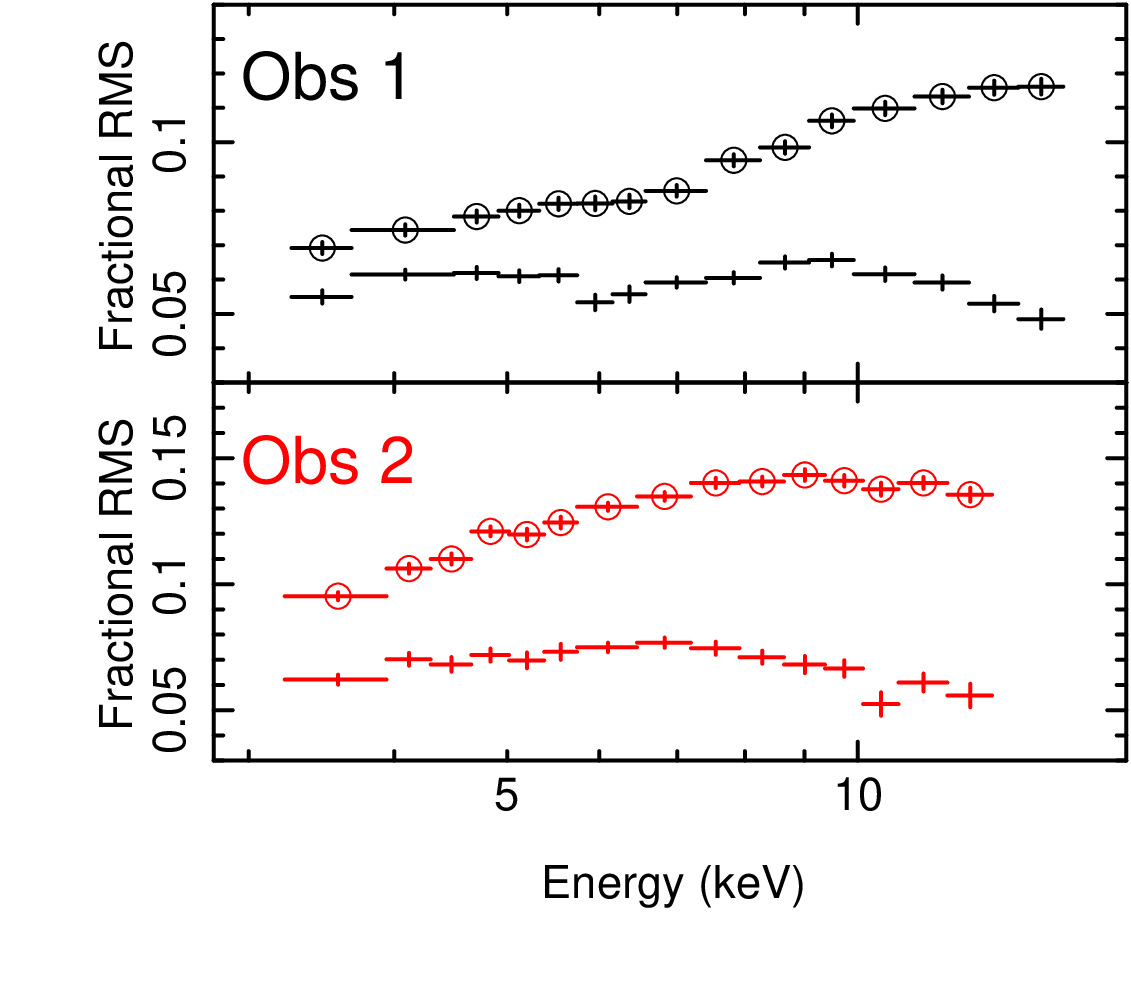}
\vspace{-1.0cm}
 \caption{Fractional rms for the $1^{\rm st}$ (circled points) and
   $2^{\rm nd}$ (no marker) QPO harmonics as a function of energy for
   both observations.}
 \label{fig:rms}
\end{figure}

Now that we can reconstruct a waveform for the full band, we can
reconstruct a waveform for each energy channel by generalising
equation \ref{eqn:w} to
\begin{equation}
w(E,\phi) = \mu(E) \left\{ 1 + \sqrt{2}\sum_{j=1}^{J} \langle
  \sigma_j(E) \rangle \cos[ j\phi - \Phi_j(E) ] \right\}.
\label{eqn:we}
\end{equation}
For each energy channel, we extract a light curve for which it
is again simple to measure the mean. We fit a multi-Lorentzian model
to the power spectrum of each light curve and define the rms in the
$1^{\rm st}$ and $2^{\rm nd}$ QPO harmonics as the integral of the
corresponding Lorentzian function (following
e.g. \citealt{Axelsson2014}). Figure \ref{fig:rms} shows the measured
fractional rms in the $1^{\rm st}$ (circles) and $2^{\rm nd}$ (points)
harmonics as a function of channel energy.

\begin{figure}
 \includegraphics[height=8.0cm,width=8.5cm,trim=0.0cm 0.0cm 0.0cm
 0.0cm,clip=true]{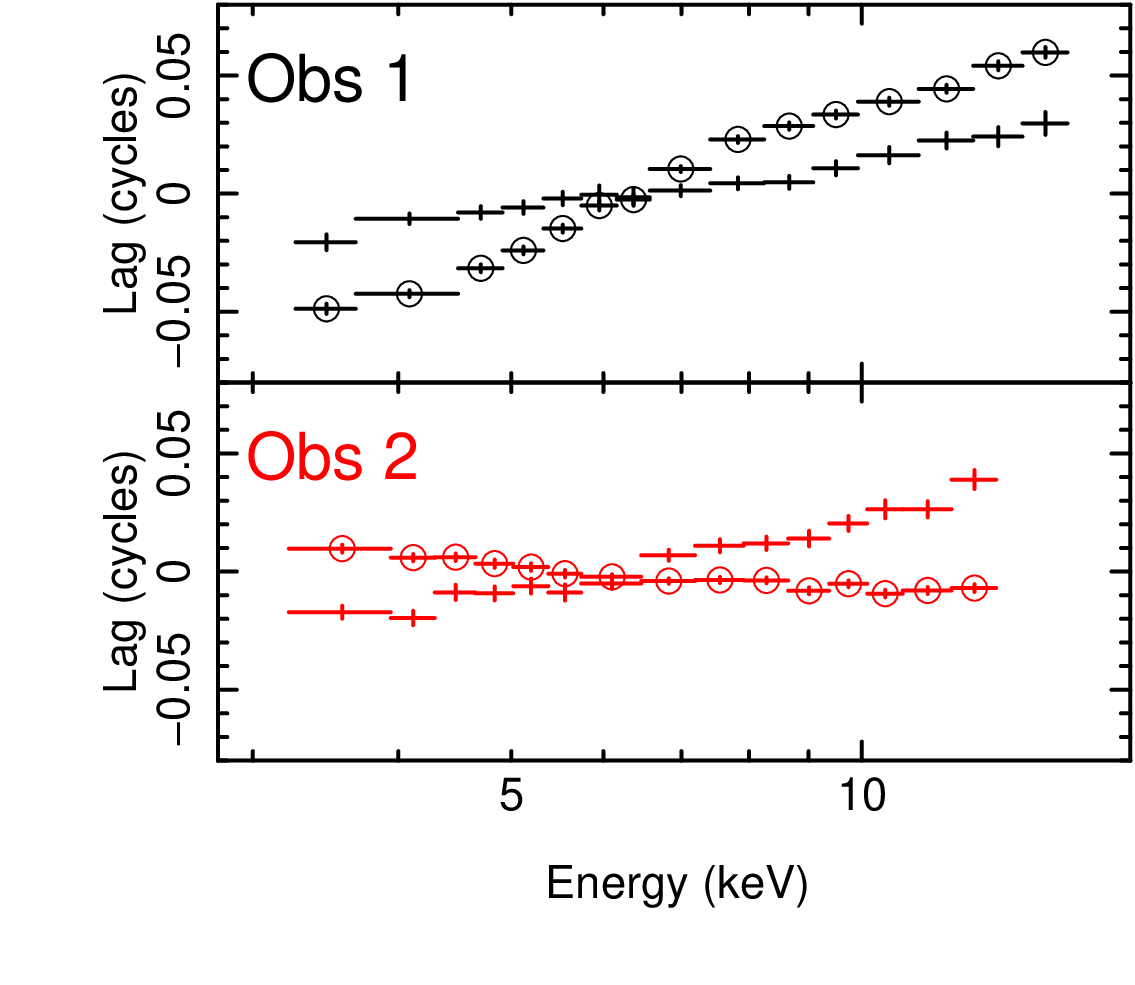}
\vspace{-1.0cm}
 \caption{Phase lags between a given energy channel and the full band
   for the $1^{\rm st}$ (circled points) and $2^{\rm nd}$ (no marker)
   QPO harmonics.} 
 \label{fig:lags}
\end{figure}

The most obvious way of measuring the phase offsets would perhaps be
to measure $\langle \psi \rangle$ for each energy channel using the
method described in section \ref{sec:phase}. We would then need to
measure the phase difference between energy bands of the first
harmonic. Instead, we maximise signal to noise by measuring the phase
lag at each harmonic, $\Delta_j(E)$, between each energy band, $E$,
and the full band. With our measure of $\langle \psi \rangle$ for the
full band, we can calculate the phase offsets using the formulae
\begin{eqnarray}
\Phi_1(E) &=& \pi / 2 + \Delta_1(E) \nonumber \\
\Phi_2(E) &=& 2( \Phi_1(E) + \langle \psi \rangle ) + \Delta_2(E).
\label{eqn:Phij}
\end{eqnarray}
We calculate the lags in the usual way by taking the cross spectrum
between each subject band, $s(E,t)$, and the reference band $r(t)$
(e.g. \citealt{vanderKlis1987}), which we define as the full band with
the subject band subtracted to avoid correlating $s(E,t)$ with
itself (\citealt{Uttley2014}). For each harmonic, we evaluate the complex cross spectrum,
$C_j(E)$, at the nearest Fourier frequency to the centroid frequency
of that harmonic. The phase lag is then
\begin{equation}
\tan\Delta_j(E) = \frac{\Im C_j(E)}{\Re C_j(E)}.
\end{equation}
Figure \ref{fig:lags} shows the lags as a function of energy for the
$1^{\rm st}$ (circles) and $2^{\rm nd}$ (points) harmonics.

\begin{figure*}
 \includegraphics[height=8.0cm,width=8.5cm,trim=0.0cm 0.0cm 0.0cm
 0.0cm,clip=true]{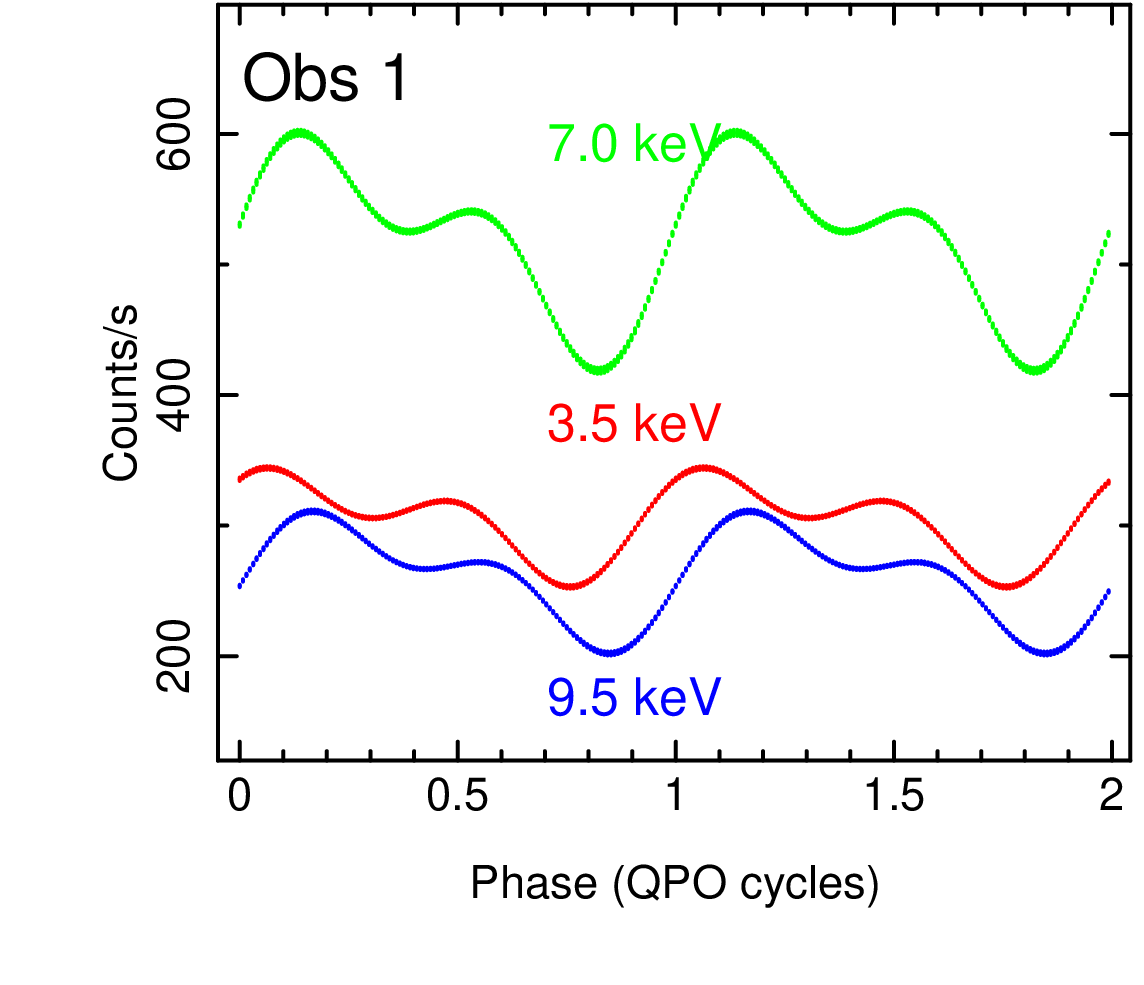}
 \includegraphics[height=8.0cm,width=8.5cm,trim=0.0cm 0.0cm 0.0cm
 0.0cm,clip=true]{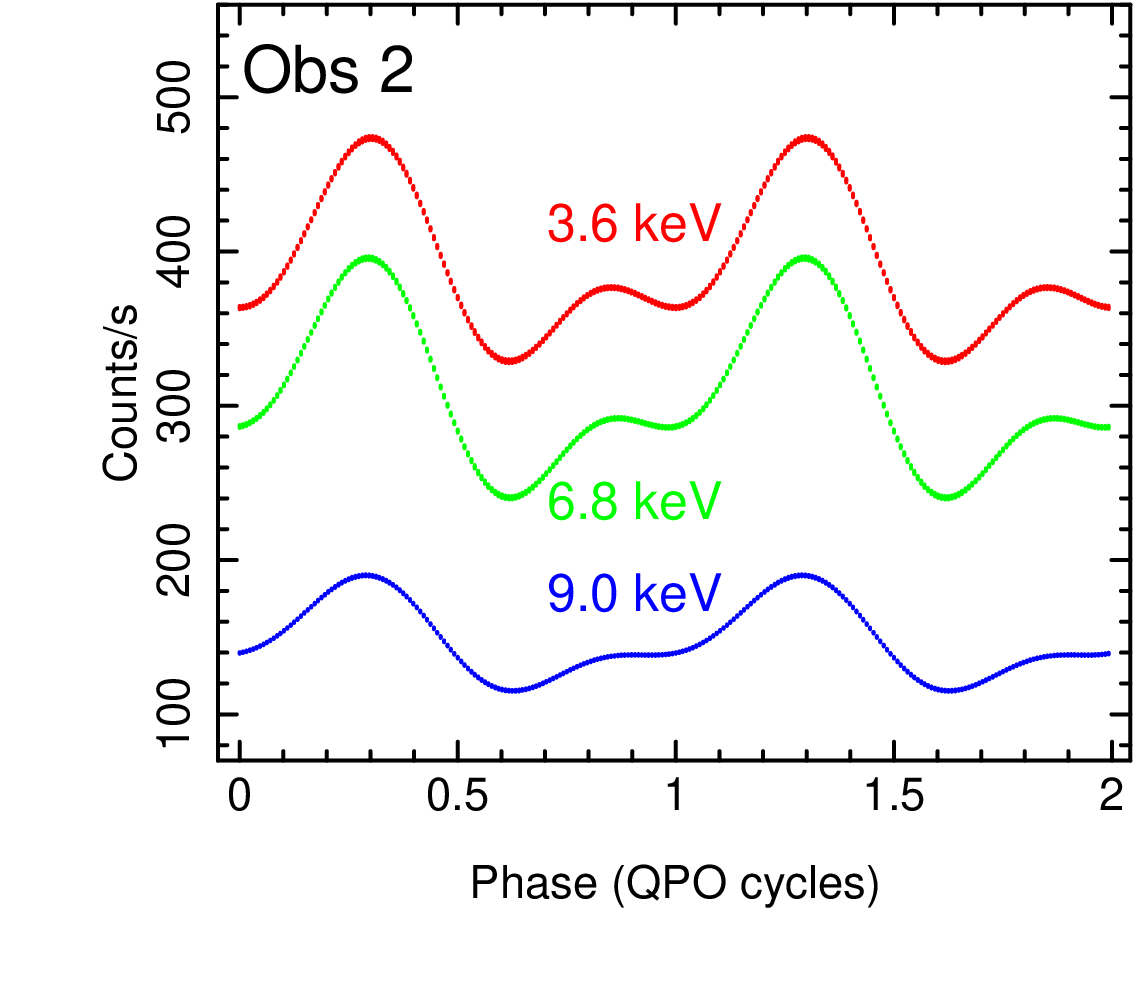}
\vspace{-0.5cm}
 \caption{Waveforms reconstructed using Equation \ref{eqn:we} for
   three different energy channels. The energy at the centre of the
   channel is labeled.}
 \label{fig:lcs}
\end{figure*}

We now have all the information required to reconstruct a waveform for
each energy channel using equation \ref{eqn:we}. Note from equations
\ref{eqn:Phij} that, even though we have only measured the phase
difference between harmonics, $\langle \psi \rangle$, in the full
band, the waveforms in different channels are free to have different
shapes. This is because their phase offsets depend on the phase lags
between energy bands which in general can be different for different
harmonics. Light curves for three channels (with the energy at the
centre of the channel labeled) are shown in Figure \ref{fig:lcs}. We
see that the waveform shape changes with energy channel for both
observations considered here. We note that, much like the case of the
full band considered in the previous section, the relation between the
waveform measured for each energy channel and the physical QPO
mechanism can potentially be biased by highly non-linear decohering
effects. If the nature of these effects is strongly energy dependent,
this could bias the energy dependence of the measured
waveforms. Again, in the absence of a full understanding of the
de-cohering mechanism, we define the waveform in each energy channel
as a periodic function with the average QPO properties.

\begin{figure*}
 \includegraphics[height=8.0cm,width=8.5cm,trim=0.0cm 0.0cm 0.0cm
 0.0cm,clip=true]{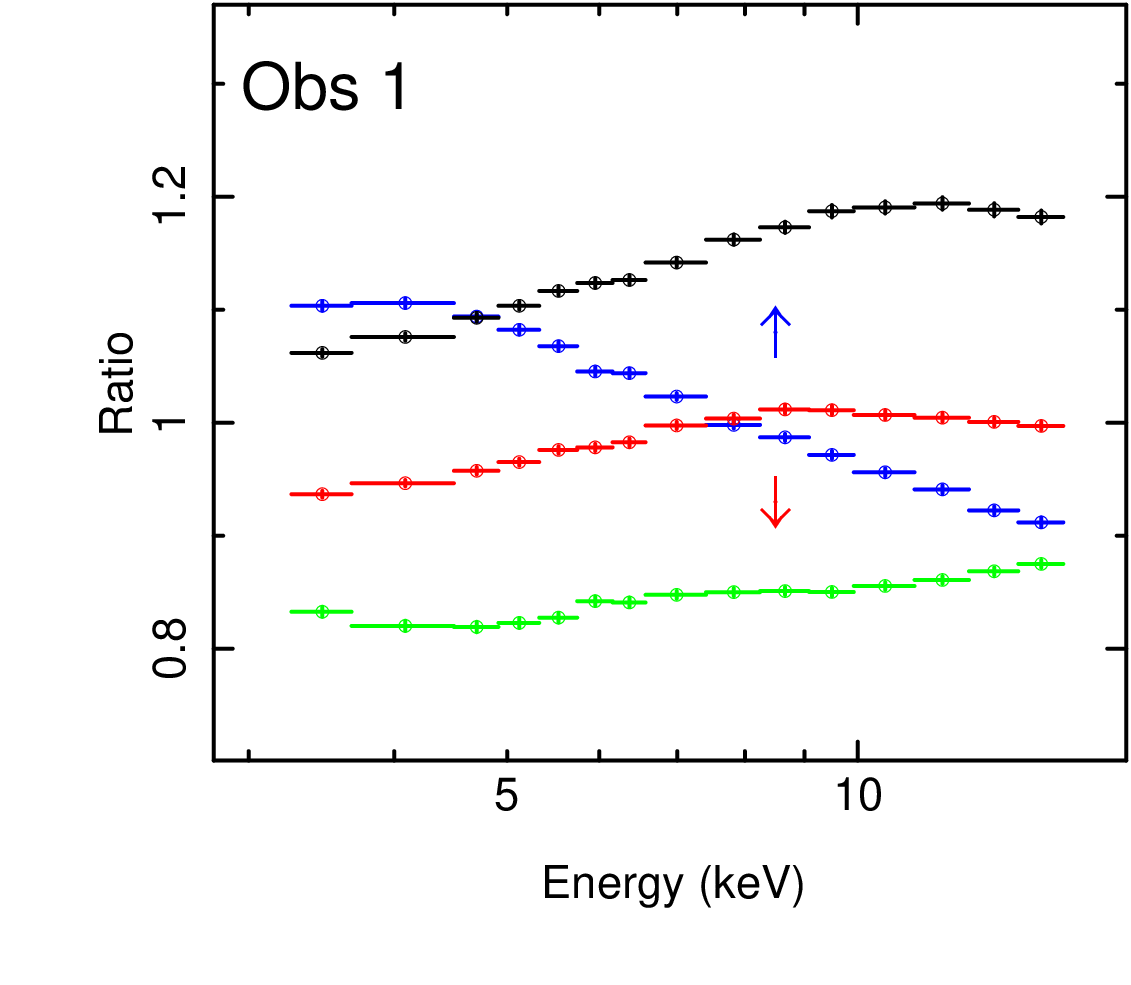}
 \includegraphics[height=8.0cm,width=8.5cm,trim=0.0cm 0.0cm 0.0cm
 0.0cm,clip=true]{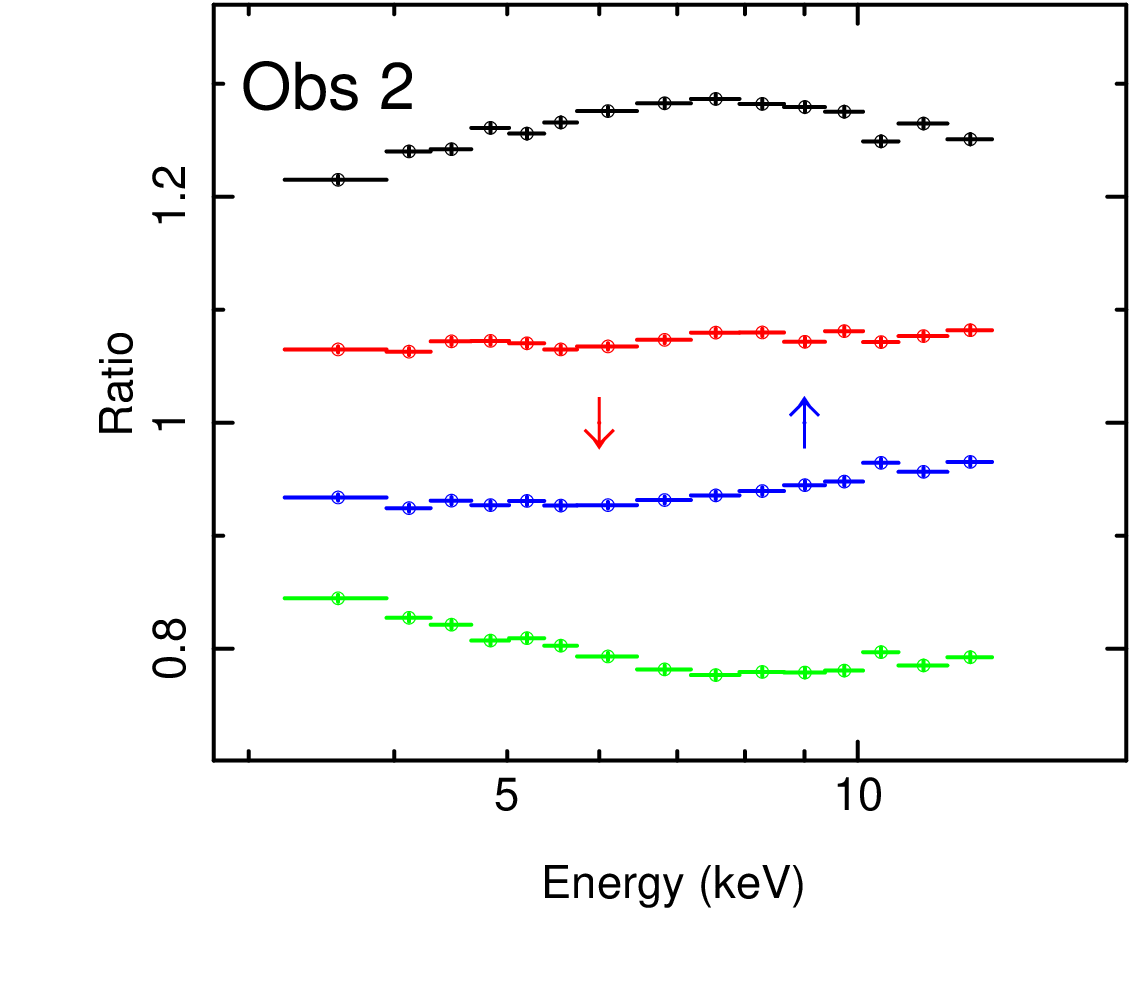}
\vspace{-0.5cm}
 \caption{The spectrum corresponding to 4 QPO phases, plotted as a
   ratio to the mean spectrum. Phases are selected to be
   representative of rising (blue), peak (black), falling (red) and
   trough (green) intervals. For observation 1, these phases are
   $\phi=$ 0, 0.1875, 0.625 and 0.75 QPO cycles respectively. For
   observation 2, they are $\phi=$ 0, 0.3125, 0.4375 and 0.625
   cycles. For both observations, we clearly see spectral pivoting.}
 \label{fig:ratio}
\end{figure*}

We calculate an error estimate for the waveform in each channel in the
same way as we did for the full band: we calculate a waveform for each
segment and measure the dispersion around $w(E,\phi)$. This involves
the additional step of calculating the phase lags $\Delta_j(E)$ for
each segment. We can now plot the count rate as a function of energy
for any given number of QPO phases: i.e. we can plot and analyse QPO
phase resolved spectra. In Figure \ref{fig:ratio}, we plot spectra for
4 QPO phases, represented as a ratio to the phase averaged
spectrum. For both observations, we see strong spectral
pivoting. Even though different values of QPO phase are not
statistically independent, it is important to note that different
energy channels \textit{are} statistically independent. This means
that we can use $\chi^2$ statistics in order to fit models to the
spectrum for each phase and study how spectral parameters vary with
QPO phase. We note that \cite{Miller2005} studied the phase resolved
behaviour of `Type-C' QPOs in GRS 1915+105 (in fact, they studied our
observation 2) by selecting spectra from high and low flux
intervals. Our technique takes this further, allowing us to study the
evolution of spectral parameters with QPO phase rather than for just
two phases. Phase resolved spectroscopy has also been used to
investigate the `Heartbeat' state of GRS 1915+105
(\citealt{Neilsen2011}) and also the QPO in the Active Galactic
Nucleus RE J1034+396 (\citealt{Maitra2010}), although we note that all
previous analyses have assumed the phase difference between harmonics
to be constant, in contrast to this Paper.

\section{Spectral Modelling}
\label{sec:spectral}

We use \textsc{xspec} version 12.8 to fit the spectral model
\begin{equation}
\textsc{phabs}~*~\textsc{smedge}~*~(\textsc{ewgaus}~*~\textsc{nthComp} ),
\end{equation}
for 16 QPO phases. Here, \textsc{phabs} accounts for interstellar
absorption for a given hydrogen column density $N_h$ and a given set
of elemental abundances. We fix $N_h$ to a reasonable value consistent
with previous analyses of these observations
(e.g. \citealt{Miller2005}) and assume the solar abundances of
\cite{Wilms2000}. The model \textsc{nthComp} (\citealt{Zdziarski1996};
\citealt{Zycki1999}) calculates a Comptonisation spectrum consisting
of a power law (photon index $\Gamma$) between low and high energy
breaks, governed respectively by the seed photon and 
electron temperature, $kT_{bb}$ and $kT_{e}$. Since the data do not
extend beyond $15$ keV, we cannot constrain the electron temperature
so arbitrarily fix $kT_{e}=100$ keV. In contrast, we allow $\Gamma$
and $kT_{bb}$ to go free in the fit. The model \textsc{smedge} mimics
the shape of a smeared reflection edge in the PCA bandpass and has
input parameters $E_{\rm Edge}$, $f$ and $W$, which govern the
position, depth and width of the reflection edge. In our fits, we fix
these parameters to reasonable values. We find that the data do not
statistically require a disc component due to the high column density
surrounding GRS 1915+105 and the hard response of the PCA.

Since the equivalent width (EW) of the iron line is of interest, we
define a new \textsc{xspec} model \textsc{ewgaus}, which is simply a
Gaussian function with three parameters: centroid energy in keV
($E_c$) width in keV ($\sigma$) and EW in eV. Thus, the only
difference to the standard \textsc{xspec} Gaussian function is that
the EW is an input parameter rather than the line flux. We define this
as a convolution model, since we must determine from the continuum the
normalisation required to give the line the specified EW, for which we
use Brent's method. Note that, even though this is defined as a
convolution model, this is \textit{not} the mathematical operation: we
simply add the Gaussian to the continuum, we define a convolution
model purely to allow the continuum to be input to the model.

In the following subsection we present the results of our spectral
fits for both observations. We allow 5 parameters of physical interest
to be free in the fit as a function of QPO phase: the continuum
parameters $\Gamma$ and $kT_{bb}$, plus the iron line parameters
$E_c$, $\sigma$ and EW. For each of these parameters, we use an
\textit{f-test} comparing a fit with the parameter held constant to
the best fit model to asses if it varies with QPO phase, and with what
statistical significance.

\begin{figure*}
 \includegraphics[height=12.0cm,width=8.7cm,trim=0.0cm 0.0cm 0.0cm
 0.0cm,clip=true]{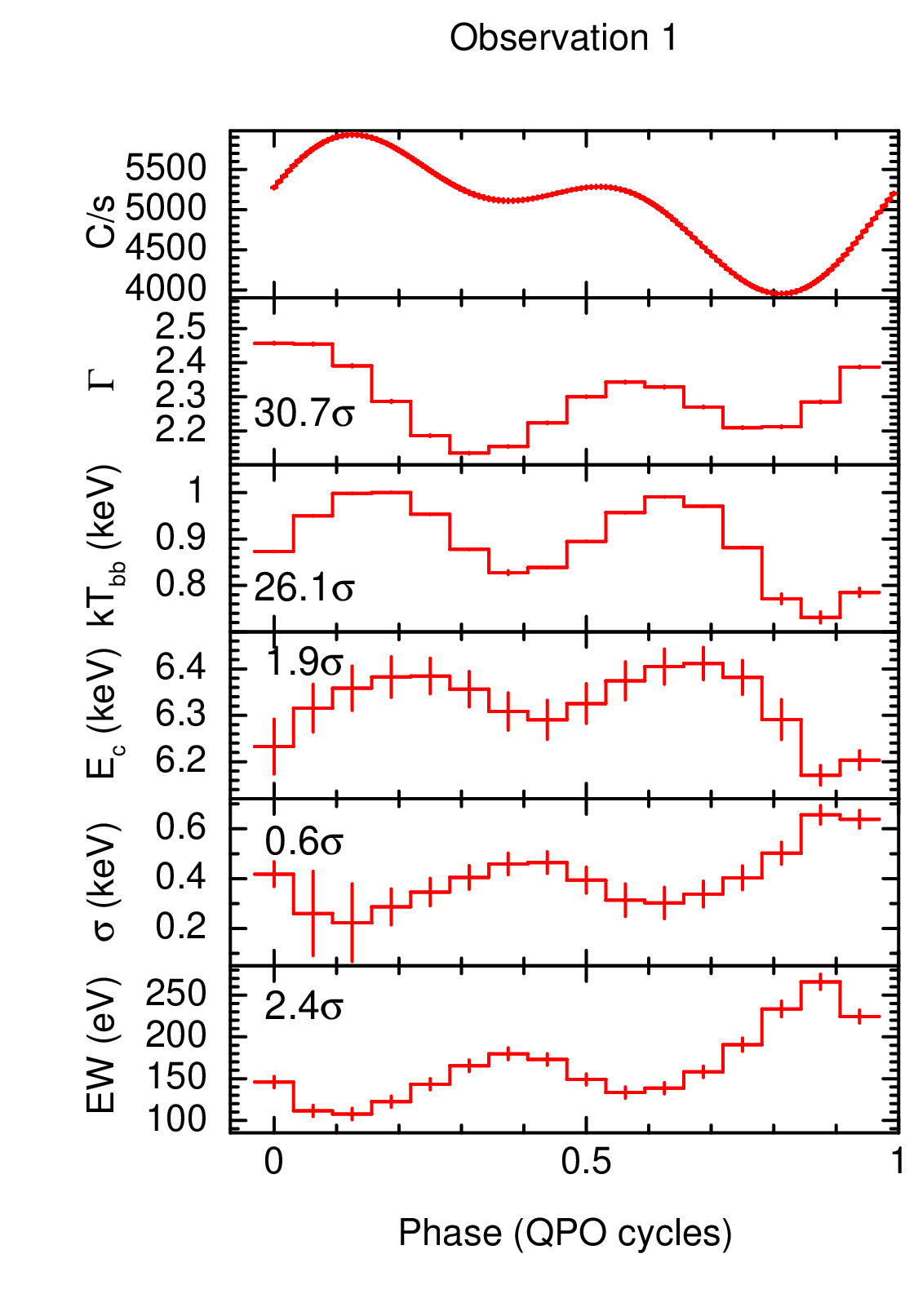}
 \includegraphics[height=12.0cm,width=8.7cm,trim=0.0cm 0.0cm 0.0cm
 0.0cm,clip=true]{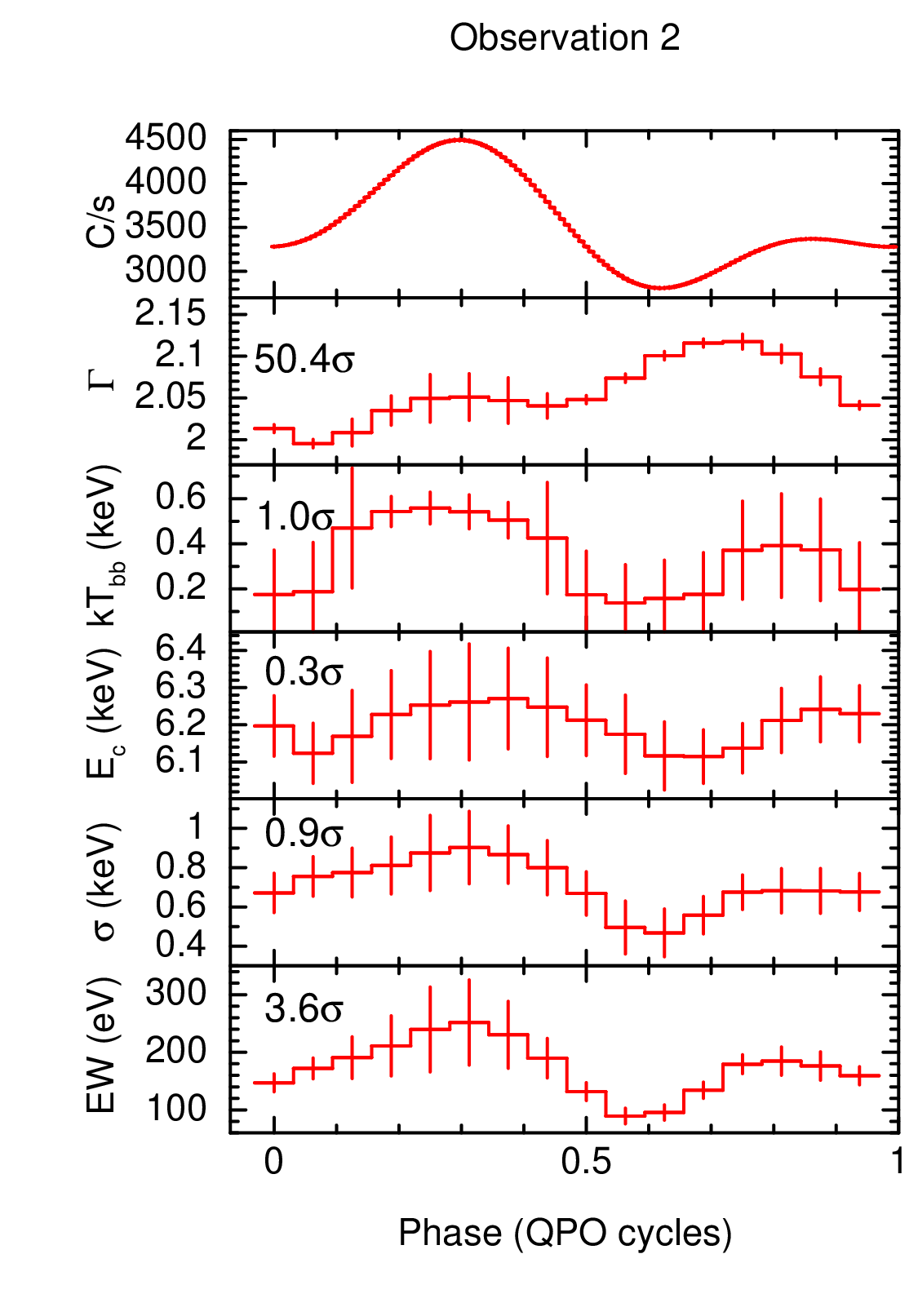}
\vspace{-0.5cm}
 \caption{Best fit parameters of our spectral fit plotted as a
   function of QPO phase. Both observations show highly significant
   spectral pivoting and we see a modulation in the relative strength
   of the iron line. The statistical significance of each modulation
   is quoted in the corresponding plot.}
 \label{fig:paras}
\end{figure*}

\subsection{Results}
\label{sec:results}

\subsubsection{Observation 1}

We achieve a good fit with $\chi^2_\nu=158.07/144$ by freezing the
hydrogen column density to $N_{\rm h}=5.2\times 10^{22}~{\rm cm}^{-2}$
and the \textsc{smedge} parameters to $E_{\rm Edge}=8.25$ keV, $f=0.3$
and $W=5$ keV. Figure \ref{fig:paras} (left) shows the evolution of the 5
physically interesting parameters across 16 phases, with the full band
waveform also reproduced at the top for reference (error bars are all
$1\sigma$). We quote with what statistical significance each spectral
parameter varies with QPO phase in the top-left corner of each
panel. We see that $\Gamma$ and $kT_{bb}$ vary with a very high
significance and both contain a strong $2^{\rm nd}$ harmonic. We also
see that the iron line parameters vary, also with a strong $2^{\rm
  nd}$ harmonic, but non above the 3$\sigma$ level. Clearly, the
points as plotted are incompatible with a constant but, when each iron
line parameter is held constant for an alternative fit, changes in
other parameters can, to some extent compensate. Note that the
systematic nature of these modulations does not alone indicate they are
real. A random $1 \sigma$ fluctuation in, say, the measured rms can
potentially result in a systematic looking modulation in the phase
resolved spectrum. We must therefore use the f-tests to asses
significance.

\subsubsection{Observation 2}

We again achieve a good fit with $\chi^2_\nu=144.06/144$, this
time by freezing the hydrogen column density to $N_{\rm h}=5.4\times
10^{22}~{\rm cm}^{-2}$ and the \textsc{smedge} parameters to $E_{\rm
  Edge}=8.5$ keV, $f=0.3$ and $W=5$ keV. We plot the QPO phase
evolution of the best fit parameters on the right of Figure
\ref{fig:paras}. We again see a highly significant modulation of
$\Gamma$ but this time $kT_{bb}$ only varies with $1\sigma$
confidence. The modulations in the iron line centroid and width are
not statistically significant, but the EW varies with $3.6 \sigma$
confidence. Our results are consistent with those of
\cite{Miller2005}, who analysed spectra for this observation selected
for high and low flux intervals.

\subsection{Interpretation}
\label{sec:interp}

We can conclude with high statistical confidence that the spectral
index varies with QPO phase for both observations and also that the
parameter $kT_{bb}$ varies with phase in observation 1. We also find a
$> 3\sigma$ modulation in the iron line EW for observation 2. Here, we
discuss possible interpretations of these modulations as well as
speculating about what modulations we may expect to see in the iron
line shape for higher quality data sets.

\subsubsection{Continuum parameters}

We can picture the observed spectral variability, on the simplest
level, as a power law with changing index and normalisation. If the
total flux in a broad energy band \textit{lags} the power law index,
hard photons will lag soft photons. This is because $\Gamma$ is a
proxy for spectral \textit{softness}, and thus this means the peak
flux lags the softest spectrum, or in other words the hardest spectrum
lags the peak flux. This corresponds to a positive gradient in the lag
vs energy spectrum. For observation 1, we do indeed find that the
total PCA count rate lags $\Gamma$ by $\approx 0.29$ cycles for the
fundamental and $\approx 0.06$ cycles for the $2^{\rm nd}$ harmonic,
which is consistent with the positive gradient seen in Figure
\ref{fig:lags} for the lag spectrum of both harmonics. For observation
2, we instead find that the total count rate \textit{leads} $\Gamma$
by $\approx 0.42$ cycles for the fundamental and \textit{lags}
$\Gamma$ by $\approx 0.12$ cycles, which is consistent with the
negative gradient of the lag spectrum of the fundamental and the
positive gradient for the $2^{\rm nd}$ harmonic. It is this spectral
pivoting which is at the heart of the generic models for alternating
phase lags recently proposed by \cite{Misra2013} and
\cite{Shaposhnikov2012}.

Physically, this spectral pivoting can either be attributed to changes
in Comptonisation or perhaps changes in the reflection hump, which we
do not model here. If the pivoting were exclusively down to a changing
flux in the reflection hump, the spectral hardness would track the
reflection fraction. We would therefore expect an anti-correlation
between $\Gamma$ and the iron line EW - i.e. a phase difference of 0.5
cycles for each harmonic. Since the phase difference for the strong
$2^{\rm nd}$ harmonic is $\sim 0.24$ and $\sim0.23$ cycles for
observation 1 and 2 respectively, it seems likely that there is at
least some pivoting of the Comptonised spectrum itself. This can
result from modulations in the temperature, $kT_e$, and/or optical
depth, $\tau$, of the corona. The simple formula $\Gamma-1\propto
1/[\tau kT_{\rm e}]$ approximates the case of thermal Compton
scattering (\citealt{Pietrini1995}). Although we cannot discern
between these two interpretations here, we note that it should be
possible to measure both the electron temperature and the shape of the
reflection hump as a function of QPO phase by carrying out a similar
analysis with \textit{the Nuclear Spectroscopic Telescope   ARray}
(\textit{NuStar}; \citealt{Harrison2013}), which has a high spectral
resolution and reasonable throughput up to $\sim 70$ keV.

The fits for observation 1 also clearly require $kT_{bb}$ to change
with QPO phase with very high significance. Since we do not detect a
direct disc component, it is difficult to interpret exactly what this
means. This could really be a measure of the seed photon
temperature. Alternatively, the \textsc{nthComp} component could be
mimicking a combination of weak direct disc emission plus Comptonised
emission. If the true disc flux were to increase in this scenario, the
low energy cut-off of the \textsc{nthComp} component would move to a
lower energy in order to find a fit. Thus a \textit{minimum} in
$kT_{bb}$ could, counter-intuitively, correspond to a \textit{maximum}
in direct disc flux. We are unable to determine if this is the case
with these data, but will investigate for less absorbed sources with a
visible direct disc component in future.

\subsubsection{Iron line parameters}

Although we see a systematic variation in iron line EW with QPO phase
for both observations, the modulation is only statistically significant
($3.6\sigma$) for observation 2. For both observations, $EW(\phi)$ has
a strong $2^{\rm nd}$ harmonic. The ratio of the amplitude in the $2^{\rm nd}$
harmonic relative to the $1^{\rm st}$ is $\sigma_2/\sigma_1 \approx 1.5$ 
and $\sigma_2/\sigma_1 \approx 1.0$ for observations 1 and 2
respectively, in contrast to $\sigma_2/\sigma_1 \approx 0.70$ and
$\sigma_2/\sigma_1 \approx 0.59$ for the total flux. Modulations in the
iron line EW indicate that the reflection fraction changes throughout
the QPO cycle. This could be because the accretion geometry is
changing over the cycle (i.e. \textit{geometric origin}) and thus the
solid angle of the emitter as seen by the reflector and/or the solid
angle of the reflector as seen by the observer is
changing. Alternatively, the geometry may remain constant and the
reflection fraction changes purely because an increase in illuminating
flux ionises the disc further, thus increasing the reflection albedo
of the reflector (e.g. \citealt{Matt1993}). In the latter case, the
change in ionisation will be very fast compared with the QPO period
and thus the modulation in EW can be modelled as
\begin{equation}
EW(\phi) \propto C(\phi)^\delta,
\end{equation}
where $C(\phi)$ is the continuum flux and $\delta>1$ is a
constant. This non-linear response model can explain why $EW(\phi)$
has a stronger $2^{\rm nd}$ harmonic than the total flux [assuming
this is a proxy for $C(\phi)$]. It cannot, however, explain any phase
lag between $EW(\phi)$ and the total flux. In observation 1, the iron
line EW leads the total flux by $\sim 0.45$ cycles for both the  $1^{\rm
  st}$ two harmonics - although we caution that the EW modulation is
only $2.4\sigma$ significant. In observation 2, the EW leads the total
flux by $\sim 0.044$ and $\sim 0.013$ cycles for the $1^{\rm st}$ and
$2^{\rm nd}$ harmonics respectively. Since this is compatible
with zero lag on the $10\%$ level, our results do not fully rule out
the EW modulation in observation 2 resulting purely from changes in
disc ionisation, although they strongly hint that a change in geometry
is required. Clearly, much information about the system can be learned
by carrying out this analysis on many more observations. In
particular, a significant EW modulation with a large phase lag
relative to the total count rate would provide confirmation of a
geometric QPO origin.

In addition, we see very tentative hints of modulations in the
centroid and width of the iron line, which are not significant enough
to make conclusions. Nonetheless, the prospect of detecting shifts in
the iron line shape in future is exciting since the line profile is
heavily influenced by Doppler shifts from rapid Keplerian rotation
close to the BH, as well as general relativistic effects
(e.g. \citealt{Fabian1989}). In the model of \cite{Ingram2009}, the
QPO results from Lense-Thirring precession of the entire inner
accretion flow. This model predicts that the iron line should rock
between red and blue shift as the inner flow preferentially
illuminates respectively the receding and approaching sides of the
disc (\citealt{Ingram2012a}). Although we do not have the statistics
to test this prediction here, we note that the observed iron line
centroid and width can only realistically be influenced by dynamical
smearing (i.e. variable Doppler and gravitational shifts) or
ionisation. As discussed above, increased illuminating flux will
further ionise the disc material. In addition to changing the albedo,
this will also increase the rest frame energy and width of the iron
K$_\alpha$ line (e.g. \citealt{Matt1993}). Thus the geometry may be
fixed but varying degrees of ionisation cause modulations in $E_c$ and
$\sigma$. In this case, however, $E_c$ and $\sigma$ must both be in
phase with the illuminating flux. Thus, observing the centroid to vary
out of phase with the width would provide strong evidence of a
non-azimuthally symmetric QPO mechanism.

\section{Discussion \& Conclusions}
\label{sec:discussion}

We present a QPO phase resolved spectral analysis for 2
observations of GRS 1915+105. In order to do this, we have developed a
method to reconstruct QPO waveforms in each energy channel from the
average properties of the first two QPO harmonics. We note that our
method does not \textit{a priory} assume that there is a well defined
average underlying waveform, rather we independently verify that this
is the case for the two observations considered. We determine the
distribution of phase differences, $\psi$, between QPO harmonics over
many short segments of time and formally demonstrate that $\psi$
varies tightly around some mean value, $\langle \psi \rangle$. This
indicates that the QPO is not simply an uncorrelated noise process with excess
variability at harmonically related frequencies, but instead has a
well defined underlying waveform. This conclusion can be
inferred \textit{a posteriori} from the bicoherence measurements of
\cite{Maccarone2011}. We measure the mean phase difference to be
$\langle \psi \rangle / \pi = 0.667  \pm   9.9\times 10^{-3}$ and
$\langle \psi \rangle / \pi = 0.133   \pm   4.6\times 10^{-3}$ for
observations 1 ($\nu_{\rm   qpo}=0.46$ Hz) and 2 ($\nu_{\rm
  qpo}=2.26$ Hz) respectively. Clearly, the phase difference evolved
between these two observations. Since these observations display very
different QPO frequencies, it is possible that $\langle \psi \rangle$
correlates in some way with QPO frequency. To test this in upcoming
work, we will measure $\langle \psi \rangle$ for many more
observations.

We reconstruct an estimate for the underlying waveform from these
measurements of $\langle \psi \rangle$ and the rms variability in each
harmonic. This now opens up the possibility of using waveform fitting
to test theoretical QPO models (e.g. \citealt{Veledina2013}), in
direct analogy to the pulse profile modelling technique routinely used
for coherent NS pulses (e.g. \citealt{Poutanen2003}). Reconstructing a
waveform in each energy channel allows us to constrain spectra for 16
QPO phases which we fit with a model including Comptonisation and
reflection, with the latter accounted for simply by Gaussian and
smeared edge components. We find that the photon index of Comptonisation varies
with very high significance for both observations but the modulation
in best-fit seed photon temperature is only statistically significant
for observation 1. We conclude that the
former could be due to some combination of changes in the electron
temperature or optical depth of the corona and changes in the
amplitude of the reflection hump in the spectrum. This degeneracy can
be broken by carrying out a similar analysis up to high energies, as
is now possible with \textit{NuSTAR}. As for the seed photon
temperature, this is difficult to interpret since we do not include a
direct disc component in our model due to the high absorption column
around GRS 1915+105 and the hard response of the PCA. More light can
be shed on this result by studying sources with a lower absorption
column in states with more prominent direct disc emission, preferably
with \textit{XMM Newton} which has a softer response than
\textit{RXTE}.

Our best fit model shows a modulation in the EW of the Gaussian
representing the iron line, which has a significance of $2.4\sigma$
and $3.6\sigma$ for observations 1 and 2 respectively. This indicates
that the reflection fraction varies over the QPO cycle, which in turn
implies that the accretion geometry is changing over the QPO cycle. We
note, however, that our results can possibly be explained with a
constant accretion geometry with the iron line EW variations given by
changes in ionisation state of the disc material. This interpretation
seems fairly unlikely however, especially since the QPO amplitude
appears to correlate with the source inclination angle
(\citealt{Heil2014a}; \citealt{Motta2014b}). Phase resolved spectral
analysis of more observations may well soon provide the required body
of evidence to conclude that the QPO does indeed have a geometric
origin. 

We also see tentative hints that the iron line shape may change with
QPO phase, but do not achieve the required statistics to make a
conclusion. Modulations in the iron line shape have been predicted
for a few QPO models (\citealt{Karas2001}; \citealt{Tsang2013};
\citealt{Ingram2012a}), all due to variable Doppler shifts. In the
precessing inner flow model, the iron line is predicted to rock
between red and blue shift as the inner flow illuminates respectively
the receding and approaching sides of the accretion disc
(\citealt{Ingram2012a}). This model predicts an anti-correlation
between the line centroid and width, since the line is dominated by
the narrow blue horn when approaching disc material is illuminated but
includes strong contributions from both the red wing and the blue horn
when the receding disc material is illuminated. In contrast, variable
disc ionisation would cause a correlation between iron line centroid
and width. These predictions can perhaps be tested by analysing more
observations, however it is clear that high quality observations with
good spectral resolution are required, as would be provided by, for
example, \textit{XMM Newton}, \textit{NuSTAR} or, best of all,
\textit{the Large Observatory For x-ray Timing}
(\citealt{Feroci2012}), should it fly.

\section*{Acknowledgments}

AI acknowledges support from the Netherlands Organization for
Scientific Research (NWO) Veni Fellowship. This research has made use
of data obtained through the High Energy Astrophysics Science Archive
Research Center Online Service, provided by the NASA/Goddard Space
Flight Center. AI acknowledges useful conversations with Lucy Heil,
Diego Altamirano, Chris Done, Phil Uttley and Tom Maccarone. We
acknowledge the anonymous referee for useful comments.

\bibliographystyle{/Users/adamingram/Dropbox/bibmaster/mn2e}
\bibliography{/Users/adamingram/Dropbox/bibmaster/biblio}

\label{lastpage}

\end{document}